\documentclass[apj,a4paper,12pt,useAMS]{emulateapj} 
\usepackage{amsmath} 

\slugcomment{Accepted by ApJ: Dec.23, 2008} 

\newcommand{\mpla}{Mod.Phys.Lett.A}
\def\lsim{~\rlap{$<$}{\lower 1.0ex\hbox{$\sim$}}}

\shorttitle{AMiBA Observation, Data Analysis and Results for SZE}
\shortauthors{Wu {\it et al.}}

\begin{document}

\title{AMiBA Observations, Data Analysis and Results for Sunyaev-Zel'dovich Effects}

\author{
Jiun-Huei~Proty~Wu\altaffilmark{1,2},
Paul~T.~P.~Ho\altaffilmark{3,4}, Chih-Wei~Locutus~Huang\altaffilmark{1,2},
Patrick~M.~Koch\altaffilmark{3}, Yu-Wei~Liao\altaffilmark{1,2},
Kai-Yang~Lin\altaffilmark{3,1}, Guo-Chin~Liu\altaffilmark{3,5}, 
Sandor~M.~Molnar\altaffilmark{3}, Hiroaki~Nishioka\altaffilmark{3},
Keiichi~Umetsu\altaffilmark{2,3}, Fu-Cheng~Wang\altaffilmark{1,2},
Pablo~Altamirano\altaffilmark{3}, Mark~Birkinshaw\altaffilmark{6},
Chia-Hao~Chang\altaffilmark{3}, Shu-Hao~Chang\altaffilmark{3},
Su-Wei~Chang\altaffilmark{3}, Ming-Tang~Chen\altaffilmark{3},
Tzihong~Chiueh\altaffilmark{1,2}, Chih-Chiang~Han\altaffilmark{3},
Yau-De~Huang\altaffilmark{3}, Yuh-Jing~Hwang\altaffilmark{3}, 
Homin~Jiang\altaffilmark{3}, Michael~Kesteven\altaffilmark{7},
Derek~Y.~Kubo\altaffilmark{3}, Katy~Lancaster\altaffilmark{6},
Chao-Te~Li\altaffilmark{3}, Pierre~Martin-Cocher\altaffilmark{3},
Peter~Oshiro\altaffilmark{3}, Philippe~Raffin\altaffilmark{3},
Tashun~Wei\altaffilmark{3}, Warwick~Wilson\altaffilmark{7}}

\altaffiltext{1}{Department of Physics, Institute of Astrophysics, \& Center for Theoretical Sciences,
	 National Taiwan University, Taipei 10617, Taiwan}
\altaffiltext{2}{LeCosPA Center, National Taiwan University, Taipei 10617, Taiwan}
\altaffiltext{3}{Institute of Astronomy and Astrophysics, Academia Sinica, P.~O.~Box 23-141, Taipei 10617, Taiwan}
\altaffiltext{4}{Harvard-Smithsonian Center for Astrophysics, 60 Garden Street, Cambridge, MA 02138, USA}
\altaffiltext{5}{Department of Physics, Tamkang University, 251-37 Tamsui, Taipei County, Taiwan}
\altaffiltext{6}{University of Bristol, Tyndall Avenue, Bristol BS8 1TL, UK}
\altaffiltext{7}{Australia Telescope National Facility, P.O.Box 76, Epping NSW 1710, Australia}

\begin{abstract}

We present observations, analysis and results
for the first-year operation of AMiBA, 
an interferometric experiment designed to study cosmology 
via the measurement of Cosmic Microwave Background (CMB).
AMiBA is the first CMB interferometer operating at 3~mm
to have reported successful results,
currently with seven close-packed antennas of 60-cm diameter
giving a synthesized resolution of around $6^\prime$. 
During 2007 AMiBA detected the Sunyaev-Zel'dovich effects (SZE)
of six galaxy clusters at redshift $0.091 \leq z \leq 0.322$.
An observing strategy with on-off-source switching is used 
to minimize the effects from electronic offset and ground pickup. 
Planets were used to test the observational capability of AMiBA and 
to calibrate the conversion from correlator time-lag data to visibilities. 
The detailed formalism for data analysis is given.
We summarize our early tests including observations of planets and quasars, and 
present images, visibility profiles, 
the estimated central coordinates, sizes, and SZE amplitudes of the galaxy clusters.
Scientific implications are summarized.
We also discuss possible systematic effects in the results.

\end{abstract}



\section{Introduction}	
The Cosmic Microwave Background (CMB) has been used as a window
through which to study not only the early Universe but also its
evolutionary history. 
The challenge of such observations arises from
the fact that the CMB carries information spanning a period of about
14 billion years, so that specially-designed instruments and analysis methods 
are required to separate the effects originated from different physical processes. 
One recent focus is the measurement of the Sunyaev-Zel'dovich effects (SZE)
resulting from the hot gas in galaxy clusters \citep{s-z,birkinshaw-lancaster-07}. 
Several CMB experiments are dedicated to this purpose, such as 
ACT \citep{Kosowsky2003}, AMI \citep{Kneissl2001}, APEX-SZ \citep{apex-sz08},
OVRO/BIMA-SZE \citep{bima-sz06},
SPT \citep{Ruhl2004}, SZA \citep{sza07}, and OCRA-p \citep{ocra07}.
Here we report the first results of SZE observations
with the Y.~T.~Lee Array for Microwave Background Anisotropy \citep[AMiBA;][]{amiba07-ho}.

AMiBA is devoted to CMB observation with 
particular emphasis on detecting SZE clusters,
the CMB anisotropy, and possibly cosmic defects
\citep{sz04-lin,amiba04-umetsu,amiba04-wu,amiba07-ho-mpla,amiba07-wu-mpla}.
The instrument is described in \citet{amiba07-ho}, \citet{amiba07-chen}, and
\citet{amiba07-koch-mount}.
In our initial configuration with seven close-packed antennas of 60-cm diameter,
we concentrated on pointed observations of the SZE and the measurement 
of the CMB temperature power spectrum \citep{amiba07-ho-mpla,amiba07-wu-mpla}. 

At the AMiBA operating frequency (86-102 GHz; $\sim 3$-mm wavelength) the SZE is
expected to induce a decrement in CMB intensity as compared with the
undistorted Planckian spectrum (see Sec.~\ref{sec-tel-sze}). This provides a powerful tool for
discriminating between galaxy clusters and other astronomical sources, 
because the latter normally emit photons characteristic of much harder
spectra than the CMB, and therefore always induce an increment in
intensity rather than a decrement. By measuring the SZE intensity
deficit and its profile, we hope to probe not only the cluster physics
but also the related cosmic origins.

This paper is organized as follows. 
In Section~2 we describe the AMiBA telescope
and its relation to the SZE.
In Section~3,
we summarize our initial observations in 2007
with emphasis on the observing strategy and target selection.
In Section~4,
we describe our analysis formalism and procedures for 
obtaining the calibrated visibilities from the raw time-lag data,
and the SZE images and profiles from the calibrated visibilities.
The main results are also presented here.
In Section~5
we discuss possible systematic errors.
In Section~6,
we summarize the scientific implications of these results,
including the estimation of Hubble constant \citep{amiba07-koch},
the scaling relationship between the SZE and X-ary derived properties \citep{amiba07-huang},
and the baryonic fraction when our SZE data are jointly analyzed with the Subaru lensing data \citep{amiba07-umetsu}.
Brief conclusions are given in Section~7.
We have other companion papers investigating the system performance \citep{amiba07-lin}, 
the foreground \citep{amiba07-liu}, and data integrity \citep{amiba07-nishioka}.

\section{AMiBA Telescope and SZE}
\label{sec-tel-sze}

AMiBA is an interferometric experiment initiated in Taiwan in 2000 
and dedicated on Mauna-Loa (3400~m in elevation), Big Island, Hawaii on October 3, 2006.
It has dual-channel receivers operating at 86--102 GHz, 
designed to have full polarization capabilities \citep{amiba07-chen}.
Currently it has seven close-packed Cassegrain antennas of 60-cm diameter \citep{koch06}
giving a synthesized resolution of 6 arcminutes (see Sec.~\ref{sec-res}),
expandable to a total of 19 elements with a synthesized resolution of about 2 arcminutes.
The project has been funded for an expansion to 13 elements with dishes of 1.2-m diameter. 
The 13-element system is expected to start operating in the early 2009.
Its capability in studying the SZE science is investigated by \citet{amiba07-molnar}.

The receiver-antenna elements are reconfigurable and co-mounted on a six-meter platform,
which is driven by a hexapod mount \citep{amiba07-koch-mount}.
Each element has a cooled heterodyne receiver, consisting of HEMT amplifiers of 46 dB in amplification, 
subharmonic mixers, and 2--18 GHz IF amplifiers. 
For each baseline, the signals from two dual-channel receivers are cross-correlated 
in an analogue 4-lag correlator, 
whose time-lag outputs are convertible to two complex visibilities 
corresponding to the upper and lower frequency bands (see Sec.~\ref{formalism}).
The cross correlation between the L and R polarization modes (the dual channels) of a pair of receivers 
enables the measurement of the four Stokes parameters, $T$, $Q$, $U$, and $V$.
Currently AMiBA operates with only two cross-polarization modes of LL and RR,
focusing on the measurement of $T$.
The typical receiver noise temperatures are between 80 and 110~K \citep{amiba07-lin}.

In 2007, the seven-element AMiBA focuses on targeted SZE observations.
The theoretically expected SZE in flux density is
\begin{equation}
	\label{eq-sz}
	\Delta I_{\rm SZE}(f)= \Delta I_{\rm tSZE}(f,y) +  \Delta I_{\rm kSZE}(f,\tau, v_{\rm p}),
\end{equation}
where $\Delta I_{\rm tSZE}$ and $\Delta I_{\rm kSZE}$ are the thermal and kinematic SZE respectively, defined as
\begin{eqnarray}
	\label{eq-sz-2}
  \Delta I_{\rm tSZE}(f,y) & = & I_0 y \left[g(x)+\delta_{\rm rel}(f,T_{\rm e})\right], \\
	\Delta I_{\rm kSZE}(f,\tau, v_{\rm p}) & = & - I_0 \beta\tau h(x).
\end{eqnarray}
Here
$f$ is the observing frequency,
$I_0 = 2(kT_{\rm CMB})^3/(hc)^2\approx 2.7\times 10^8$~Jy~Sr$^{-1}$
for a CMB temperature of $T_{\rm CMB}=2.725K$,
$y = k\sigma_{\rm T}/(m_{\rm e} c^2)\int T_{\rm e} n_{\rm e} dl$ is the Compton parameter,
$g(x) = h(x) [x/\tanh (x/2)-4]$ with $x = hf/(k T_{\rm CMB})$,
$\delta_{\rm rel}$ is a relativistic correction,
$\beta = v_{\rm p}/c$ is the peculiar radial velocity in units of speed of light,
$\tau = \sigma_{\rm T} \int n_{\rm e} dl$ is the optical depth,
$h(x) = x^4 e^x/(e^x-1)^2$,
$c$ is the speed of light, $m_{\rm e}$ is the electron mass,
$n_{\rm e}$ is the electron number density in the cluster,
$T_{\rm e}$ is the electron temperature,
$k$ is the Boltzmann constant,
$h$ is the Planck constant,
and
$\sigma_{\rm T}$ is the Thomson cross section.

For the thermal effects,
the $\Delta I_{\rm tSZE}$ has a maximum decrement
with respect to the CMB Planckian spectrum $I_{\rm CMB}(f) = I_0 x^3 / (e^x-1)$
at $f\approx 100$~GHz.
This optimal frequency for SZE observations is well covered 
by the AMiBA operating frequency of $f=[86,102]$~GHz.
For AMiBA $g(f)\approx -3.4$ at the center frequency $f=94$~GHz,
with a variation of less than $\pm 10\%$ for the range of $[86,102]$~GHz.
Therefore for a typical massive cluster with $y\sim 10^{-4}$,
$T_{\rm e}\sim 10$~keV, $\tau\sim y m_{\rm e} c^2 / (k T_{\rm e}) \approx 5\times 10^{-3}$,
$v_{\rm p} = 1000$~km$/$s, and an angular size of $\theta=6^\prime$,
we expect the SZE signals to be $\Delta I_{\rm tSZE} \pi (\theta/2)^2\sim -200$~mJy
and $|\Delta I_{\rm kSZE}| \pi (\theta/2)^2 \sim 20$~mJy for AMiBA,
indicating $|\Delta I_{\rm kSZE}/\Delta I_{\rm tSZE}|\approx 10$~\%.
The fractional contribution from the relativistic correction $\delta_{\rm rel}$ in such case is about 7~\%.
Thus we expect the AMiBA SZE signals to be dominated by the thermal effects,
with a feature of decrement in flux density.
We also note that
$|I_{\rm CMB}(94\textrm{GHz})|\approx 2.9\times 10^{8}$~Jy~Sr$^{-1}$,
so that $|\Delta I_{\rm tSZE}/I_{\rm CMB}|\approx 3\times 10^{-4}$.

The operating frequency of $[86,102]$~GHz also has the advantage
that the SZE signals are less affected by the point sources
when compared with the observations at a lower frequency,
which is more typical for interferometric experiments
such as AMI \citep[15 GHz; ][]{Kneissl2001}, SZA \citep[27--35 GHz; ][]{sza07},
and OVRO/BIMA \citep[30 GHz; ][]{Reese2002}.
This is because typically the point sources have power-law spectra of flux density
with negative spectral indices, so that they contribute less towards higher frequencies.
We note that
although SZA also has instruments at 90--98~GHz,
their baselines are about ten times longer than ours
providing information at much smaller angular scales 
that is complementary to what AMiBA can supply.
Another advantage of the AMiBA frequency range is that
it also suppresses the Galactic synchrotron radiation and dust emission.
Detailed investigation about the AMiBA foreground is presented in \cite{amiba07-liu}.
These features in frequency and resolution make AMiBA a unique CMB telescope,
complementing the capability of other CMB projects.

\section{Observations}
\label{sec-obs}

\subsection{Primary Beams}
\label{primary-beam}

\begin{figure}
	\plotone{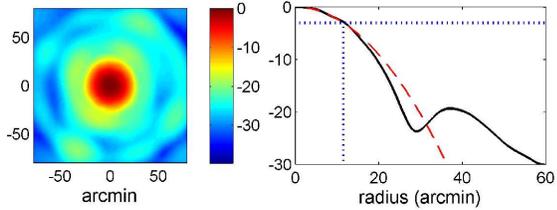}
  \caption{\small 
  (a) Left: the measured primary beam of one of the 60-cm antennas.
  (b) Right: azimuthally averaged beam shape (solid) and a Gaussian fit (dashed);
  the dotted lines indicate the FWHM.
  In both plots the intensity is in units of db normalized to the peak value. 
  \label{fig-beam}}
\end{figure}

Before the 60-cm antennas were mounted onto the platform,
their beam patterns were individually measured in the far field ($\sim 100$~m)
by scanning a wide-band white source with a system retrofitted from a commercial equatorial mount \citep{koch06}.
Figure~\ref{fig-beam}(a) shows an example.
An azimuthal average of the beam pattern (Figure~\ref{fig-beam}(b))
reveals a Full-Width-at-Half-Maximum (FWHM) of about 23 arcminutes.
The first side lobe is about 20 db below the primary peak and the beam profile is as per designed.
The index of asymmetry \citep{asym-beam} is calculated showing that
the discrepancy of the beam pattern from its azimuthal average is below 10\% within the FWHM.
All beam patterns of the antennas are well fitted by a Gaussian beam of 23 arcminutes 
within the 2-$\sigma$ region for $<10\%$ error.

\subsection{u-v Coverage and Resolution}
\label{sec-res}

\begin{figure}
	\plotone{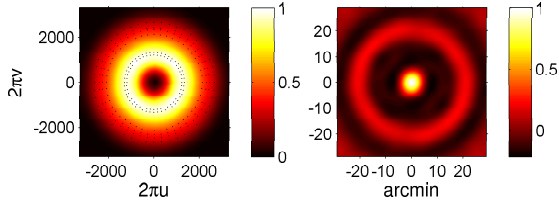}
  \caption{\small  
  The $u$-$v$ coverage of a typical observation by seven-element AMiBA (left) and
  its corresponding noise-weighted synthesized beam (the dirty beam; right).
  In the left plot, the color scale indicates the relative sensitivity
  in $u$-$v$ space, while the dots mark the $u$-$v$ modes sampled by the
  typical AMiBA observation.
  \label{fig-uv_coverage}}
\end{figure}

In 2007, AMiBA was used with the seven 57.6-cm diameter antennas
close-packed on the platform, providing 21 baselines of three different lengths
$L^b=60.6$~cm, 105.0~cm, and 121.2~cm. 
Each baseline operates at two frequency channels centered at $f=90$~and 98~GHz, each with a band width of 8~GHz.
Thus the three baseline lengths correspond to six multipole bands
$\ell=2\pi\sqrt{u^2+v^2}=2\pi L^b f/c=[1092,1193]$, $[1193,1295]$, $[1891,2067]$, $[2067,2243]$, $[2184,2387]$, $[2387,2590]$.
These baselines have a six-fold symmetry \citep{amiba07-ho}.
To achieve better $u$-$v$ coverage and thus a better imaging capability,
we observe each target at eight different polarization angles of the platform
with an angular step of 7.5 degrees, uniformly covering azimuthal angles in the $u$-$v$ plane 
with 48~discrete samples for each of the three different baseline lengths at each of the two frequency bands.
This makes a total of 288 discrete $u$-$v$ samples.
Figure~\ref{fig-uv_coverage} shows such typical $u$-$v$ coverage,
with the corresponding noise-weighted point spread function from a typical observation, 
the so-called `dirty beam'. 
The FWHM of the dirty beam is about 6~arcminutes, 
defining the synthesized resolution of the seven-element AMiBA.
When using an equal-weighted (rather than the current noise-weighted) scheme to construct the dirty beam, 
we still obtain a similar resolution of about 6~arcminutes
because the noise level for each baseline is about the same.

\subsection{Observing Strategy}

To minimize, if not completely remove, the effects from ground
pickup and from a DC component in the electronics that leaks into the
data, we adopt a `two-patch' observing strategy that alternates between
the target and a region of nominally blank sky.  
This strategy is similar to that used by CBI for the measurement of CMB polarization \citep{cbi-04}.

In this mode, a leading patch that contains our target is tracked for three minutes, 
and then a trailing patch of the presumably blank sky is tracked for the same period of time. 
The trailing patch is located at an RA $\rm 3^m10^s$ greater than the RA of the target.
Since it takes 10~seconds for the telescope to slew between the fields, 
the two patches are observed over identical azimuth and elevation tracks. 
During both tracks the platform polarization angle is controlled so that 
each baseline corresponds to a fixed $u$-$v$ mode. 
The recorded data during the two tracks should then contain the same contribution from ground pickup,
which is removed by a subsequent differencing of the two tracks.
This strategy requires that both the electronic offset and the ground emission are stable within 6 minutes, 
which is far shorter as compared with the measured time dependence \citep{amiba07-lin}.
The penalty for adopting such a two-patch strategy is the loss of efficiency by a factor of $4$, 
where a factor of 2 comes from doubling the observing time 
and the other factor of 2 from doubling the noise variance when differencing the two patches.
We also note that there is a CMB component as well as possible combinations from 
Galactic foreground and extragalactic point sources in the differencing map
though at a level much lower than the expected SZE of massive clusters \citep{amiba07-liu}.
Nevertheless before observations we did check our fields for such contaminations.

\begin{figure}
	\plotone{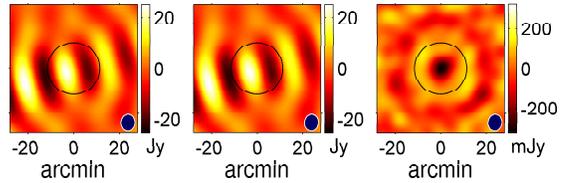}
  \caption{\small
  Modulation between the target (left) and the blank sky (center)
  successfully removes the strong contributions from ground pickup and
  electronic offset. The difference map (right) clearly reveals the
  signal from cluster A2142.
  \label{fig-2-patch}}
\end{figure}

Figure~\ref{fig-2-patch} demonstrates the application of the two-patch observing strategy for SZE mapping.
The left panel is an image constructed from the leading patch, containing cluster Abell~2142
(see Sec.~\ref{formalism} for the formalism and procedure of data analysis).
The middle panel shows the corresponding image from the trailing patch of blank sky.
It is clear that both patches are contaminated by signals much stronger than the SZE 
(mainly from the ground pickup), and that the images are closely similar.
The difference of the two patches reveals the cluster signal (right panel),
whose signal strength is only about 1\% of the contamination signal in individual patches.

\subsection{Calibration Events}
\label{sec-cal-eve}

To calibrate these data, we observe planets roughly every 3 hours using the same 2-patch observing strategy.
This time the two patches are each observed for four minutes and separated by ${\rm 4^m 10^s}$ in RA.
Such a separation in RA is equivalent to an angular separation of at least 57 arcminutes,
assuring that leakage of the planetary signal into the second patch is small,
less than $0.1\%$ of the planetary flux density
according to the primary beam pattern measured in Sec.~\ref{primary-beam}.
The detailed formalism on how the data are converted to calibrated visibilities 
is presented in Sec.~\ref{calibration},
and the determination of the planetary flux densities is discussed by \cite{amiba07-lin}.
Long-duration observations of the planets show
that the system is stable at night (between about 8pm and 8am)
so that a planetary calibration every three hours controls the calibration error 
to be within 5\% in gain and 0.1 radian in phase \citep{amiba07-lin}.

To test the overall performance of our observing, calibration, and analysis strategy, 
we first used the Sun to successfully phase-calibrate an observation of Jupiter,
and then Jupiter to phase and amplitude calibrate an observation of Saturn. 
The left plot of Figure~\ref{fig-saturn-quasar} shows a dirty image of Saturn 
calibrated by an observation of Jupiter made four hours later. 
The FWHM of this image is identical to that of the dirty beam shown in Figure~\ref{fig-uv_coverage},
and the index of asymmetry \citep{asym-beam} of the cleaned image is well below 5\%.

\subsection{2007 observations and SZE Targets}

\begin{figure}
	\plotone{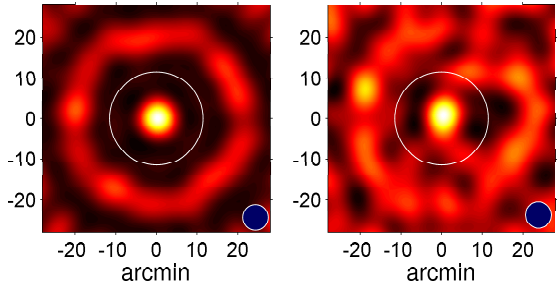}
  \caption{\small  
  Dirty images of Saturn (left) and the quasar J0325+2224 (right),
  both calibrated by Jupiter using the formalism presented in Sec.~\ref{formalism}.
  \label{fig-saturn-quasar}}
\end{figure}

After first light of the seven-element AMiBA in September 2006, 
which was a drift scan on Jupiter and whose image was quickly constructed, 
several months of start-up tests were used to study the instrumental properties, 
to fine tune our system, to study the ground pickup, and to find an optimal observing strategy. 
During this period, Saturn, Mars, Venus, and the Crab Nebula
were also observed for system tests. In February 2007, when testing
the two-patch observing strategy, we detected a first extragalactic
object, quasar J0325+2224 with a total (unresolved) flux density of
550~mJy (right plot of Figure~\ref{fig-saturn-quasar}).

After further fine tuning of the system, in April~2007 we detected our
first SZE cluster, Abell 2142 with a central SZE decrement of around
-320 mJy per beam. 
During 2007 we observed a total of six S-Z clusters with $0.091 \leq z \leq 0.322$:
A1689, A1995, A2142, A2163, A2261, and A2390.

The observations were always at night time, when the system is stable.
Due to the weather, system engineering, and observability of the targets,
a total of roughly 150 observing hours was spent 
in observing the clusters or their trailing fields over a span of 5~months,
from April to August in 2007. 
During this period a lot of efforts were dedicated to the system tests and tuning
\citep{amiba07-lin,amiba07-koch-mount}.
Table~\ref{tab-sz-obs} summarizes the pointing directions, the observing and integration
times, and the synthesized beam sizes, for each of the six clusters. 
Due to the two-patch observing strategy, which modulates between on-
and off-source positions, the total time spent during cluster
observations is double the on-cluster time listed in the table. 
The integration time on source ($t_{\rm int}$) is less than the actual observing time ($t_{\rm obs}$)
because of the data trimming that will be discussed in Sec.~\ref{time-domain-processing}.

\begin{table}
	\begin{center}
  \caption{Observing log for AMiBA SZE clusters.\label{tab-sz-obs}}
	\begin{tabular}{c | c c c c c  c c}
		\tableline\tableline
		cluster	& \multicolumn{2}{c}{pointing dir.~(J2000)}	& $t_{\rm obs}$	& $t_{\rm int}$	& syn.		\\
						&		RA										& DEC							&	(hr)					&	(hr)					&	res.		\\
		\tableline
		A1689	&	$13^{\rm h}11.49^{\rm m}$ & $-1^\circ20.47'$	&	\phantom{1}8.75	&	\phantom{1}7.11	&	$6.3'$	\\
		A1995	&	$14^{\rm h}52.84^{\rm m}$ & $58^\circ02.80'$ 	&						15.90	&	\phantom{1}5.56	&	$6.6'$	\\
		A2142	&	$15^{\rm h}58.34^{\rm m}$ & $27^\circ13.61'$	&	\phantom{1}7.05	& \phantom{1}5.18	&	$6.5'$	\\
		A2163	&	$16^{\rm h}15.57^{\rm m}$ & $-6^\circ07.43'$	&	\phantom{1}7.80	&	\phantom{1}6.49	&	$6.6'$	\\
		A2261	&	$17^{\rm h}22.46^{\rm m}$ & $32^\circ07.62'$	&	          19.15	&	\phantom{1}8.87	&	$6.4'$	\\
		A2390	&	$21^{\rm h}53.61^{\rm m}$ & $17^\circ41.71'$	&						16.05	&						11.02	&	$6.4'$	\\
		\tableline
	\end{tabular}
	\tablecomments{
    Observing log for the AMiBA observations of six
    clusters of galaxies, including the pointing directions, the
    observing ($t_{\rm obs}$) and integration ($t_{\rm int}$) times per baseline,
    and the synthesized resolution
    (syn.~res.; the FWHM of the azimuthally averaged dirty beam).
  }
	\end{center}
\end{table}

The six clusters were chosen based on observational convenience, 
the expected flux density based on SZE observations at other frequencies, 
and the existence of extensive published X-ray data. 
The observational convenience requires that the targets can be seen 
at the Mauna-Loa latitude of $19.54^\circ$~N,
at the night time when the system is stable (between about 8pm and 8am),
and 
during the main observing period between April and August in 2007.
Several cluster catalogs are compiled and studied for the expected SZE flux at AMiBA frequency,
including OVRO \citep[30 GHz; ][]{ovro01}, OVRO/BIMA \citep[30 GHz; ][]{bima-ovro01,Reese2002}, 
VSA \citep[34 GHz; ][]{vsa05}, and SuZIE~II \citep[145, 221, and 355 GHz; ][]{suzie204}.
The redshift range of $z\lsim 0.3$ is so chosen that
massive clusters of a typical virial radius of few Mpc is resolvable by AMiBA,
whose resolution ($6^\prime$) corresponds to a physical scale of $1.6$~Mpc at $z=0.3$
in a flat $\Lambda$CDM cosmology with matter density parameter $\Omega_{\rm m}= 0.3$, 
cosmological constant density parameter $\Omega_\Lambda= 0.7$, and Hubble constant $H_0=70~{\rm km}\,{\rm s}^{-1}{\rm Mpc}^{-1}$.
Among these six clusters, only A1689 and A2390 are known to have
relaxed with a nearly isothermal profile.
A1689 has a regular morphology in X-ray while A2390 is elongated.
A2163, by contrast, is a merger with shocks and high-temperature regions, 
and A2142 is a merger with a strong frontal structure.

Due to the missing-flux problem,
most studies for the scientific implications of our results
require the assistance of X-ray derived properties of the clusters. 
In Table~\ref{tab-x-properties}, we summarize the critical X-ray
derived parameters for the AMiBA clusters,
which will be used by our companion science papers
\citep{amiba07-huang,amiba07-koch,amiba07-liu,amiba07-umetsu}. 
The X-ray temperatures are from 
\citet{Reese2002} for A2163, A2261, A1689, and A1995, \citet{markevitch98} for A2142, and \citet{Allen2000} for A2390.
The power-law index $\beta$ in the isothermal $\beta$-model and the core radius $\theta_{c}$ are from 
\citet{Reese2002} for A2163, A2261, A1689, and A1995, \citet{Sanderson2003} for A2142, and \citet{cite-a2390-beta} for A2390. 
The X-ray core radii are mostly below one arcminute except for A2142 and A2163.
These parameters are mostly X-emission weighted but not corrected for any cooling flows,
which we further discuss in some science analyses such as the estimation of the Hubble constant $H_0$ \citep{amiba07-koch}.

\begin{table}
	\begin{center}
	\caption{Parameters of AMiBA clusters derived from X-ray observations.\label{tab-x-properties}}
	\begin{tabular}{c | c c c c}
		\tableline\tableline
		Cluster	& $z$	& $kT_{\rm X}$							& $\beta$				& $\theta_{\rm c}$\\
					&				&	(keV)											&								&	(arcsec)				\\
		\tableline
		A1689	&	0.183	&	$9.66\pm^{0.22}_{0.20}$  	&	$0.609\pm^{0.005}_{0.005}$	&	$ 26.6\pm^{0.7}_{0.7}  $\\
		A1995	&	0.322 &	$8.59\pm^{0.86}_{0.67}$ 	&	$0.770\pm^{0.117}_{0.063}$	&	$ 38.9\pm^{6.9}_{4.3}  $\\
		A2142	&	0.091&	$9.7\pm^{1.5}_{1.1}$    	&	$0.74\pm^{0.01}_{0.01}$			&	$ 188.4\pm^{13.2}_{13.2}$\\
		A2163	&	0.202	&	$12.2\pm^{1.1}_{0.7}$	    &	$0.674\pm^{0.011}_{0.008}$	&	$ 87.5\pm^{2.5}_{2.0}  $\\
		A2261	&	0.224	&	$8.82\pm^{0.37}_{0.32}$	  &	$0.516\pm^{0.014}_{0.013}$	&	$ 15.7\pm^{1.2}_{1.1}  $\\
		A2390	&	0.232	&	$10.13\pm^{1.22}_{0.99}$	&	$0.6$		            				&	$ 28.0                 $\\
		\tableline
	\end{tabular}
	\end{center}
\end{table}

\section{Analysis Method and Results}
\label{formalism}

We describe the formalism that we use to analyze AMiBA data 
from their raw format to the calibrated visibilities, images, and cluster profiles.
AMiBA operates with a bandwidth of 16~GHz centered at 94~GHz.
Its four-lag correlators provide two pass-bands of 8~GHz processing dual linear polarizations.
The challenge for the analysis of such data arises
because the measured output of the instrument is not visibilities and
because there are only two channels of wide bands 
causing considerable band-smearing effect in frequency.
With the formalism presented here 
we successfully used the Sun to calibrate the image of Jupiter,
and then used several planets (Jupiter, Saturn, Mars, Venus) to cross-calibrate. 
Jupiter was chosen as the primary calibrator for our science results.

When we observe a general source field $S({\bf x})$ with a synthesized primary beam $B({\bf x})$,
the visibility along a baseline $b$ of length $L^b$ is
\begin{equation}
	v^b(f)= \widetilde{S}({\bf k}) \otimes \widetilde{B}({\bf k}),
\end{equation}
where 
${\bf k}$ is the corresponding $u$-$v$ mode,
$f= |{\bf k}| c / (2\pi L^b)$ is the observing frequency,
a tilde denotes the Fourier transform of a quantity,
and $\otimes$ denotes a convolution.
The four lag-outputs of a correlator for the source field can thus be modeled as
\begin{eqnarray}
	c^b(\tau_m) & = &	\Re\left\{
				\int_{f_1}^{f_2} v^b(f) R^b(f)\times \right. \nonumber\\
			&	& 
			\left.\exp\left[2\pi(f-f_0)\tau_m i+\phi^b(f) i \right]df
				\right\},
	\label{eq:vis2lag}
\end{eqnarray}
where
$\tau_m$ ($m=1\textrm{--}4$) is the time delay for each of the four lags,
$R^b(f)$ and $\phi^b(f)$ are the
instrumental  frequency-dependent gain- and phase-responses respectively,
$f_0$ is the frequency of the Local Oscillator (86 GHz),
and 
$f_1$ and $f_2$ indicate the upper and lower ends of the response frequency range
i.e.\ $[f_1,f_2]=[86,102]$ GHz for AMiBA.
Our analysis must invert Equation~(\ref{eq:vis2lag})
to obtain the calibrated visibilities $v^b(f)$.

\begin{figure*}
	\plotone{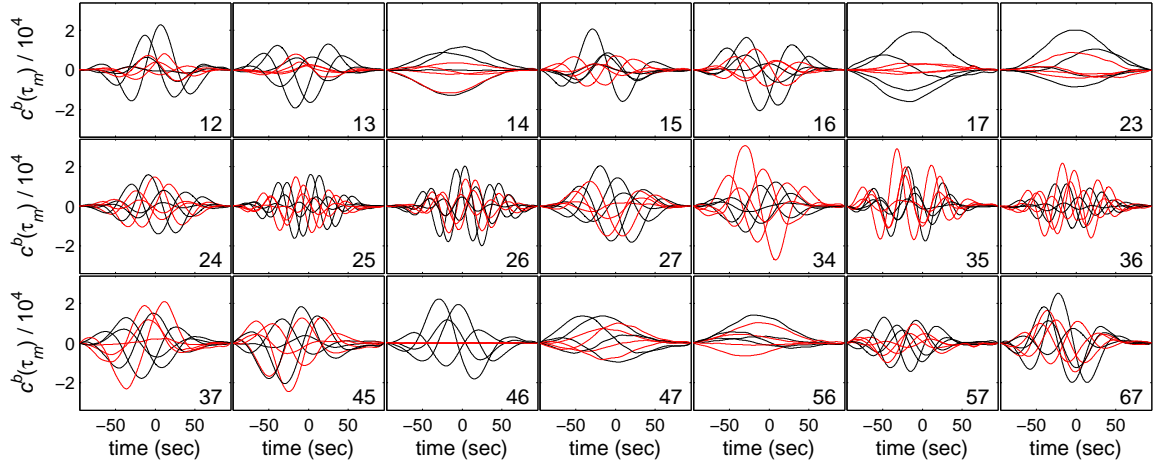}
  \caption{\small 
  First-light lag data $c^b(\tau_m)$ of a drift scan on Jupiter in September 2006.
  Each panel represents one baseline, with an `$XY$' label at the bottom-right corner
  indicating the correlation between the antennae $X$ and $Y$. 
  $X=1$ represents the central antenna, with the other six close-packed around it.
	In each panel, the four black curves and four red curves correspond to 
	the lag outputs of the LL and RR cross-polarization modes respectively	(see Sec.~\ref{sec-tel-sze}).
	Here a low-pass filtering at 0.05~Hz is used to remove high-frequency noise.
	The horizontal time axis is the offset relative to the transit time.
  \label{fig-lag}}
\end{figure*}

Figure~\ref{fig-lag} shows our first-light lag data $c^b(\tau_m)$ of a drift scan on Jupiter.
The fringing rate is determined by the angle between a baseline and the drift direction.

\subsection{Time-Domain Pre-Processing}
\label{time-domain-processing}

Initially the cluster data must be processed to remove baselines 
for which the system is malfunctioning.
In addition to checking the instrumental logs of receivers and correlators,
nightly drift-scan data for Jupiter or Saturn are used to identify obvious problems.
We then apply Kolmogorov-Smirnov (K-S) test to filter out datasets 
for which the noise appears non-Gaussian \citep{amiba07-nishioka}.
In the K-S test, a $5~\%$ significance level is used throughout and 
shown to be efficient in identifying data sets with known hardware problems.
Finally, we remove periods of data where occasional mount control problems appear.

After this flagging, we difference the tracking data $c^b(\tau_m; t)$ in the two-patch observation 
in order to remove ground pickup and the electronic DC offset.
4$\sigma$ outliers in the temporal domain are clipped before the differenced tracking data are integrated 
over the 3-minute (4-minute for the calibration) tracking period
to yield the lag data $c^b(\tau_m) \equiv \langle c^b(\tau_m; t) \rangle_t$.
Thus each two-patch observation yields a set of four lags $c^b(\tau_m)$ ($m=1\textrm{--}4$).
We carefully verified that
the noise of the lag data is white within the time scale of 10 minutes,
assuring the scaling of the noise level with the two-patch integration time \citep{amiba07-nishioka}.

Further flagging of 4$\sigma$ outliers is applied in the visibility space (see Sec.~\ref{calibration}).
Table~\ref{tab-sz-flag} shows the percentages of data that are flagged out by each step.
On average the fraction of good data is about 60\%.


\begin{table}
\begin{center}
\caption{Percentages of data flagged by each flagging step.\label{tab-sz-flag}}
\begin{tabular}{c | c c c c | c}
\tableline\tableline
Cluster & baseline	& non-Gau. & mount  & 4$\sigma$ & good \\
        & prob. 		&  noise   & prob. 	& outliers 	& data \\
\tableline
A1689	& \phantom{1}2 &						11	&  4 & \phantom{1}2 & 81 \\
A1995	&           35 &  \phantom{1}4	&  5 &           21 & 35 \\
A2142	& \phantom{1}8 & 						15	&  1 & \phantom{1}2 & 74 \\
A2163	& \phantom{1}7 &  \phantom{1}8	&  1 & \phantom{1}1 & 83 \\
A2261	&           25 &  \phantom{1}1	&  1 &           27 & 46 \\
A2390	& \phantom{1}5 & 						16	&  3 & \phantom{1}7 & 69 \\
\tableline
\end{tabular}
\end{center}
\end{table}

\subsection{Lag-to-Visibility Transform}
\label{lag-to-visibility-transform}

We must invert Equation~(\ref{eq:vis2lag}) to convert the four lags
$c^b(\tau_m)$ to two visibilities $v^b(f)$. Ideally, if $\tau_m\equiv
\tau$ is continuous, then we need only perform an inverse Fourier
transform from the $\tau$ domain to the $f$ domain. However $\tau$
here is sampled discretely, and conservation of the degrees of freedom
provided by four lag measurements implies that we can obtain at best
two uncorrelated complex visibilities, with no redundancy.

We perform the inversion first by dividing the 16-GHz frequency band into two, 
$[f_1, f_d]$ and $[f_d, f_2]$, each associated with one `band-visibility',
and define a `visibility vector' as
\begin{equation}
    {\bf v}^b
         =\left[
              \begin{array}{c}
                   \Re\{v^b([f_1,f_d])\}\\ \Im\{v^b([f_1,f_d])\}\\ \Re\{v^b([f_d,f_2])\}\\ \Im\{v^b([f_d,f_2])\}
              \end{array}
          \right].
    \label{eq:vb}
\end{equation}
If we also write the four lags in column-vector form 
${\bf c}^b \equiv c^b(\tau_m)$ ($m=1\textrm{--}4$),
then the visibility-to-lag transform resulted from Equation~(\ref{eq:vis2lag})
is simply ${\bf c}^b={\bf U}^b {\bf v}^b$,
where ${\bf U}^b$ is a $4\times 4$ matrix.
The four columns in ${\bf U}^b$ can be easily obtained from Equation~(\ref{eq:vis2lag})
as the four vectors ${\bf c}^b$
when setting each component in ${\bf v}^b$ to unity in turn while zeroing the other three components.
Subsequently the inversion is
\begin{equation}
    {\bf v}^b=\left[{\bf U}^b\right]^{-1} {\bf c}^b = {\bf V}^b {\bf c}^b.
    \label{eq:lag2vis0}
\end{equation}
Once given any lag data ${\bf c}^b$, we can use Equation~(\ref{eq:lag2vis0}) 
to construct the band-visibilities $v^b([f_1,f_d])$ and $v^b([f_d,f_2])$.

In principle, if $R^b(f)$ and $\phi^b(f)$ are accurately known, 
then ${\bf V}^b$ can be accurately determined so that there is no need for further calibration.
However, in reality $R^b(f)$ and $\phi^b(f)$ are time-dependent 
so that ${\bf V}^b$ is also time-dependent and thus needs to be calibrated for every observation. 
Because the transform~(\ref{eq:lag2vis0}) is linear, 
any reasonable choice of $f_d$, $R^b(f)$ and $\phi^b(f)$ should yield the same ${\bf v}^b$ after calibration.
Hence we make the simplest choice
$f_d=(f_1 + f_2)/2=94$~GHz, $R^b(f)=1$, and $\phi^b(f)=0$,
and then calibrate the data using the formalism described in Section~\ref{calibration}. 
This choice makes the inversion~(\ref{eq:lag2vis0}) exactly a discrete inverse Fourier transform
that accounts for the band-smearing effect.
An alternative, and equally natural, choice
might be to have equal power in the two frequency bands, with
$f_d$ so chosen that $\int_{f_1}^{f_d} R^b(f) df = \int_{f_d}^{f_2}
R^b(f) df$, but this is difficult to implement since $R^b(f)$ is
time-dependent.

\subsection{Calibration}
\label{calibration}

For calibration, we observe planets using the same 2-patch observing strategy (Sec.~\ref{sec-cal-eve}),
the same time-domain pre-processing (Sec.~\ref{time-domain-processing}), 
and the same lag-to-visibility transform (Eq.~(\ref{eq:lag2vis0}), Sec.~\ref{lag-to-visibility-transform}).
The underlying band-visibilities of these calibrators have the form 
\begin{equation}
	{\bf v}^{b({\rm thy})}_\ast=
	\left[\begin{array}{cccc}
		M_1\\ 0\\ M_2\\ 0
	\end{array} \right],
\end{equation}
where $M_1$ and $M_2$ are real constants,
corresponding to the fluxes at different frequencies.
However, the band-visibilities constructed from observational data
using Equation~(\ref{eq:lag2vis0}),
where we have chosen $R^b(f)=1$ and $\phi^b(f)=0$, will take the form
\begin{equation}
	{\bf v}^{b({\rm obs})}_\ast=\left[\begin{array}{c} 
			M_1^{b({\rm obs})} \cos\phi_1^{b({\rm obs})} \\
			M_1^{b({\rm obs})} \sin\phi_1^{b({\rm obs})} \\
			M_2^{b({\rm obs})} \cos\phi_2^{b({\rm obs})} \\
			M_2^{b({\rm obs})} \sin\phi_2^{b({\rm obs})} 
	\end{array} \right].
	\label{eq:vis-obs}
\end{equation}
Since the involved operations are linear,
there must exist a calibration matrix ${\bf C}^b$ 
that corrects this discrepancy, i.e.\
\begin{equation}
	{\bf v}^{b({\rm thy})}_{\ast} = {\bf C}^b  {\bf v}^{b({\rm obs})}_{\ast}.
\end{equation}
We may model ${\bf C}^b$ as
\begin{equation}
	{\bf C}^b = 	\left[\begin{array}{cccc} 
								a^b_{1} \cos\phi^b_{1} & -a^b_{1} \sin\phi^b_{1} & 0	& 0	\\
								a^b_{1} \sin\phi^b_{1} &  a^b_{1} \cos\phi^b_{1} & 0	& 0	\\
								0	& 0	& a^b_{2} \cos\phi^b_{2} & -a^b_{2} \sin\phi^b_{2} \\
								0	& 0	& a^b_{2} \sin\phi^b_{2} &  a^b_{2} \cos\phi^b_{2}
							\end{array} \right],
	\label{eq:Cb}
\end{equation}
where the parameters $a^b_{i}$ and $\phi^b_{i}$ account for the gain- and phase-corrections
relevant to the effects from $R^b(f)$ and $\phi^b(f)$ in Equation~(\ref{eq:vis2lag}) respectively.
By comparing the moduli and phases of $v^{b({\rm thy})}_{\ast}$ and $v^{b({\rm obs})}_{\ast}$,
we obtain 
\begin{equation}
	\begin{array}{cc}
		a^b_{1}={M_1}/{M^{b({\rm obs})}_1}, & a^b_{2}={M_2}/{M^{b({\rm obs})}_2}, \\
		\phi^b_{1}=-\phi^{b({\rm obs})}_1, & \quad \phi^b_{2}=-\phi^{b({\rm obs})}_2.
	\end{array}
\end{equation}

Finally, for a set of observed lag data ${\bf c}^b$, the calibrated band-visibility vector is
\begin{equation}
    {\bf v}^b = {\bf C}^b {\bf V}^b {\bf c}^b.
    \label{eq:lag2vis}
\end{equation}
We emphasize that this is a linear transform, 
so different initial choices for $R^b(f)$ and $\phi^b(f)$ that generate different ${\bf V}^b$
should lead to different ${\bf C}^b$ but the same final calibrated band-visibilities ${\bf v}^b$. 
For each two-patch observation, we obtain one ${\bf v}^b$ 
for each cross-polarization mode and each baseline.
For each baseline and each day,
4$\sigma$ outliers in $|{\bf v}^b|$ are flagged.

\subsection{Noise Estimation}
\label{noise-estimation}

In further analyses such as image making and model fitting for the cluster profile,
a reliable estimate for the error in the band-visibilities derived above will be needed.
First the 2-patch differenced lag data ${\bf c}^b(t) \equiv c^b (\tau_m;t)$ 
are modeled as the linear sum of the signal and noise,
${\bf c}^b(t) = {\bf s}_c^b(t)+{\bf n}^b_{c}(t)$.
Since such data for clusters are dominated by electronic noise (i.e.\ ${\bf n}^b_{c}(t) \gg {\bf s}_c^b(t)$)
and the noise is white \citep{amiba07-nishioka},
the noise variance associated with each time-integrated ${\bf c}^b \equiv \langle {\bf c}^b(t) \rangle_t$
can be directly estimated as 
\begin{equation}
	({\bf \sigma}^b_{c})^2 
	\equiv \langle \left[{\bf n}^b_{c}\right]^2 \rangle
	\equiv \langle \langle {\bf n}^b_{c}(t) \rangle_t^2 \rangle
	\approx \frac{\langle [{\bf c}^b (t)]^2 \rangle_t}{\#_t},
\end{equation}
where $\#_t$ is the number of temporal data in the time integration. 
Then $({\bf \sigma}^b_{c})^2$ constitute the diagonal elements 
of the lag-lag noise correlation matrix 
${\bf N}^b_c\equiv \langle {\bf n}^b_{c} {{\bf n}^b_{c}}^{\rm T} \rangle$.
The off-diagonal elements are found to be less than 10\%
of the diagonal elements \citep{amiba07-nishioka}
so we approximate them as zeros.

In a similar fashion, the band-visibility obtained from Equation~(\ref{eq:lag2vis}) can be modeled
as a linear sum of the signal and noise, 
${\bf v}^b (t)={\bf s}^b_v(t)+{\bf n}^b_v(t)$.
Finally the noise correlation matrix for the calibrated band-visibilities is
\begin{equation}
	\label{eq-Nv}
	{\bf N}^b_v
	\equiv \langle {\bf n}^b_v {{\bf n}^b_v}^{\rm T} \rangle
	\approx {\bf C}^b {\bf V}^b {\bf N}^b_c {{\bf V}^b}^{\rm T} {{\bf C}^b}^{\rm T}.
\end{equation}

\subsection{Image Making and Cleaning}

\begin{figure*}
	\plotone{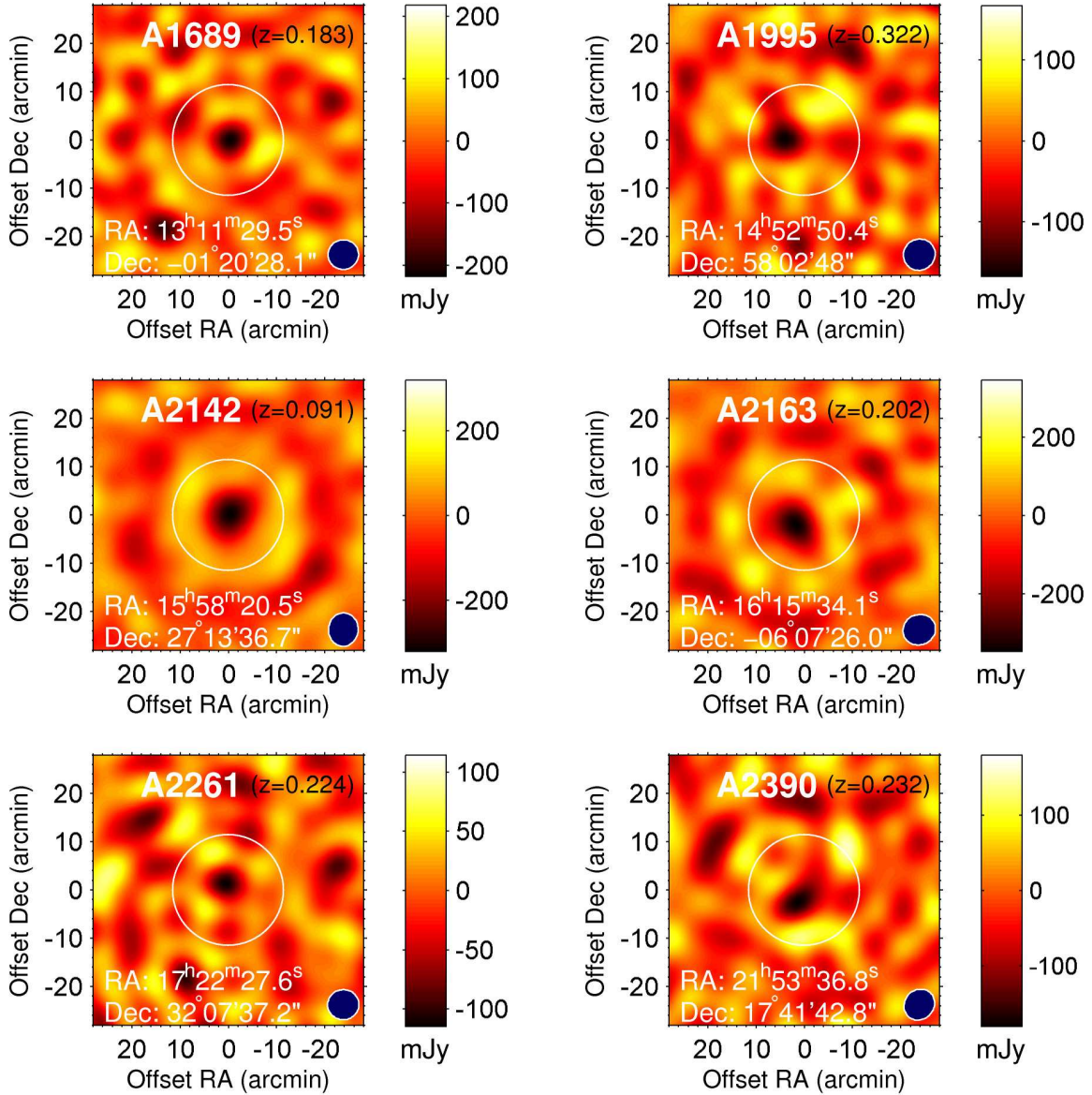}
  \caption{\small  
  Dirty images of the six SZE clusters observed by AMiBA.
  The white circles indicate the half-maximum contour of the primary beam (the field of view; $23^\prime$),
  while the blue regions at the bottom-right corners indicate the half-maximum regions 
  of the noise-weighted synthesized beams ($\sim 6^\prime$).
  The central decrements in flux density within the field of view 
  provide strong evidence for the detections of SZE clusters.
  \label{fig-sz_images}}
\end{figure*}

We construct the `dirty' image of a cluster by making a continuous inverse Fourier transform
of all the band-visibilities from all days and all baselines, 
with noise weighting based on the noise variance. 
The instantaneous $u$-$v$ coverage is improved in AMiBA operation 
by rotating the platform to eight polarization angles, giving a uniform angular interval of
$7^\circ\llap{.}5$ between sampled $u$-$v$ modes ${\bf k}$ (see Sec.~\ref{sec-res}).
Figure~\ref{fig-sz_images} shows the dirty images of these clusters. 
The flux decrement at the center of each target is evident, 
as expected for the SZE signal at the AMiBA center frequency of 94~GHz (see Sec.~\ref{sec-tel-sze}). 
The left part of Table~\ref{tab-images} summarizes the angular sizes (azimuthally averaged FWHM)
and the peak fluxes directly measured from these images.

\begin{table}
	\begin{center}
	\caption{Properties of AMiBA SZE Images\label{tab-images}}
	\begin{tabular}{c | c c | c c c c}
		\tableline\tableline
						& \multicolumn{2}{c|}{\bf dirty images}	& \multicolumn{4}{c}{\bf cleaned images}\\
		cluster	&	size		& flux	& size		& flux						& S/N 						& scale\\
						&					& (mJy)	&					& (mJy)						& ratio 					& (Mpc)\\
		\tableline
		A1689	&	($6.1'$)	& -217	&	($5.7')$	& -168						& \phantom{1}6.0	& (1.05)\\
		A1995	&	($6.8'$)	& -167	&	($6.8')$	& -161						& \phantom{1}6.4	& (1.91)\\
		A2142	&	$7.7'$		& -320	&	$9.0'$		& -316						& 13.7						& 0.92\\
		A2163	&	$7.8'$		& -347	&	$11.2'$		& -346						& 11.7						& 2.24\\
		A2261	&	($6.2'$)	& -115	&	($5.8'$)	& \phantom{1}-90	& \phantom{1}5.2	& (1.25)\\
		A2390	&	$7.4'$		& -180	&	$8.0'$		& -158						& \phantom{1}6.6	& 1.78\\
		\tableline
	\end{tabular}
\tablecomments{
Basic properties measured from the dirty and cleaned SZE images,
including the angular sizes (azimuthally averaged FWHM),
the peak flux, and the S/N ratios.
For cleaned images, the peak flux is corrected for the attenuation by the primary beam due to an offset to the pointing center.
The last column `scale' indicates the physical scale at $z$ (see Table~\ref{tab-x-properties})
corresponding to the angular size shown in the fourth column.
Here we have assumed a flat $\Lambda$CDM cosmology with
$\Omega_{\rm m}= 0.3$, $\Omega_\Lambda= 0.7$, and $H_0=70~{\rm km}\,{\rm s}^{-1}{\rm Mpc}^{-1}$.
The brackets indicate that the cluster appears unresolved in the SZE image.
}
	\end{center}
\end{table}

\begin{figure*}
	\plotone{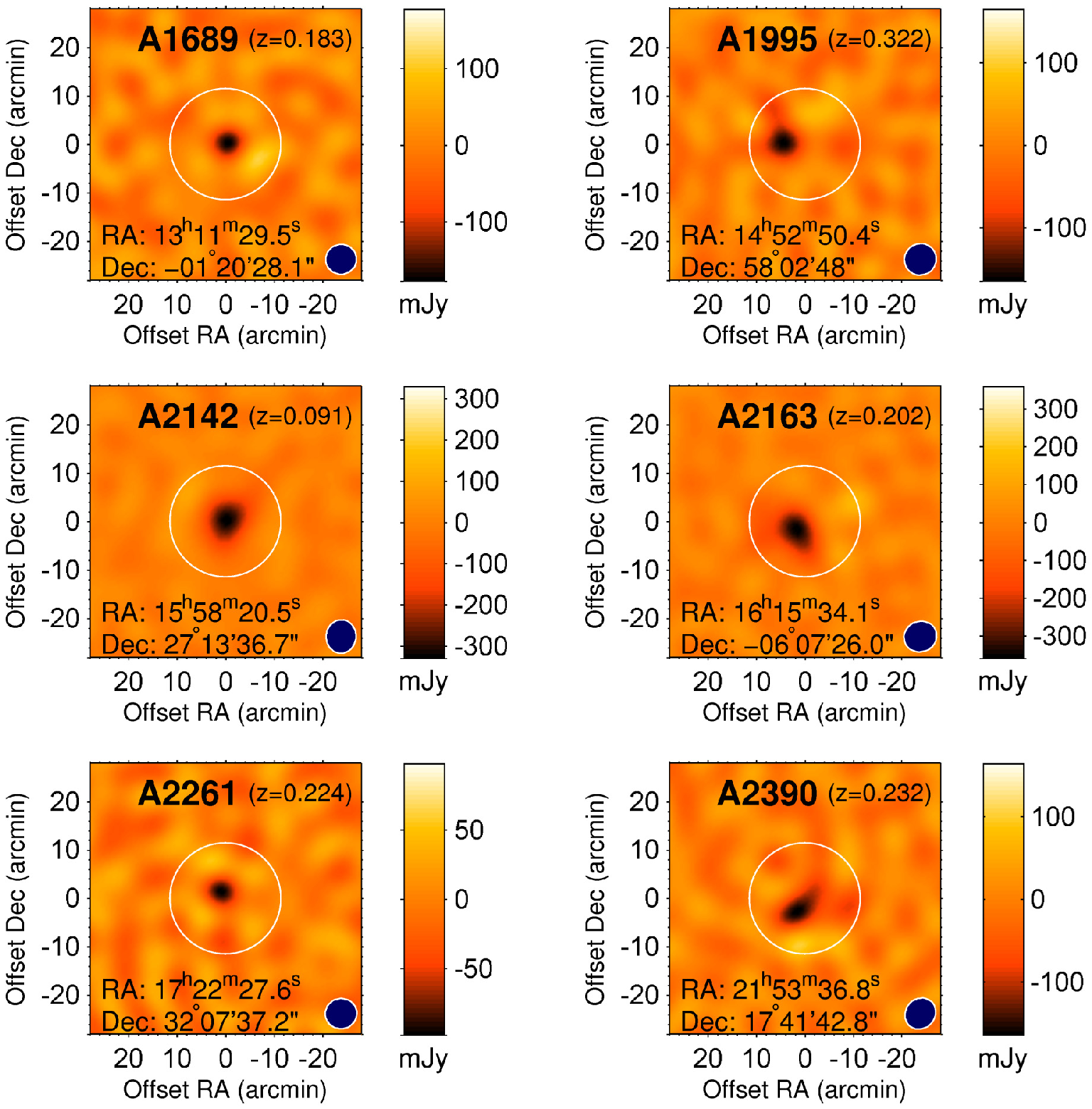}
  \caption{\small  
  Cleaned SZE images of clusters observed by seven-element AMiBA.
  The white circles indicate the half-maximum contour of the primary beam (the field of view; $23^\prime$),
  and the blue patches at the bottom-right corners show the half-maximum regions of the dirty beams ($\sim 6^\prime$).
  \label{fig-sz_images_clean}}
\end{figure*}

A dirty beam for each cluster data set is constructed using the same method, 
and the dirty image and dirty beam are then processed by
a CLEAN procedure \citep{miriad-clean} in MIRIAD \citep{miriad} to yield a cleaned image.
Figure~\ref{fig-sz_images_clean} shows the cleaned SZE images of the six AMiBA clusters, 
where the cleaned regions are indicated by the white circles, which are the FWHM contour of the primary beam.
This process significantly reduces the convolution effects from the finite $u$-$v$ coverage.
The basic properties measured from the cleaned images are summarized in the right part of Table~\ref{tab-images}.
The apparent angular sizes, as compared with the synthesized resolution of about $6^\prime$, 
indicate that we have partially resolved clusters A2142, A2163, and A2390,
while A1689, A1995, and A2261 appear unresolved in these images. 
The SZE signals observed here are at the level of few hundred mJy,
indeed consistent with the theoretical expectation $\Delta I_{\rm SZE}\sim -200$~mJy
as discussed in Sec.~\ref{sec-tel-sze}.

In Table~\ref{tab-images},
the S/N ratios are computed as the peak flux of the cleaned model 
(uncorrected for the primary-beam attenuation)
divided by the RMS of the noise residual map.
We note that the noise level in these results are dominated by the instrumental noise,
leading to the fact that
the scaling among the peak flux, S/N ratios, and integration time is consistent with the
noise equivalent flux (point-source sensitivity) of 63~mJy$\sqrt{{\rm hr.}}$ 
for the on-source integration time in a 2-patch observation \citep{amiba07-lin},
i.e.\ for a 2-hour observation, one hour per patch, the noise RMS is 63 mJy. 
Simulations also show that 
our analysis method (described in Sec.~\ref{lag-to-visibility-transform}--\ref{noise-estimation})
does not bias the SZE amplitude but induces a statistical error at the few-percent level.
The calibration error is controlled within 5~\% \citep{amiba07-lin}.
The systematic effect from CMB anisotropy is estimated to be at a similar level as the instrumental noise, 
and the point-source contamination causes an underestimate of the SZE amplitude by about 10~\% \citep{amiba07-liu}.

\subsection{Estimation of Cluster Profiles}
\label{sec-vp}

It is useful for science purposes to estimate parameters
describing the structures of the SZE clusters from the visibilities.
We employ a maximum-likelihood analysis to estimate the band-visibilities
taking the priors 
that all the visibilities represent the same circularly-symmetric source
and that the visibility moduli within a frequency band are constant.
Thus there are only six non-redundant band-visibilities $\{v_p;\; p=1\textrm{--}6\}$ 
of six multipole bands corresponding to 
the two frequency bands $[f^n_1,f^n_2]$ ($n=1,2$) of three baselines lengths
(see Sec.~\ref{sec-res} and Table~\ref{tab-mcmc}),
i.e.\ the observed band-visibilities are modeled as
\begin{eqnarray}
    \label{eq-v-vi-x0}
    v^b([f^n_1,f^n_2]| \{v_p\}, {\bf x}_0) & = &
    B({\bf x}_0-{\bf x}_1) v_{p(\in b,n)} \times \nonumber\\
    & &
    \frac{1}{k^n_2-k^n_1}
    \int_{{\bf k}^n_1}^{{\bf k}^n_2} 
    e^{i{\bf k}({\bf x}_0-{\bf x}_1)} d{\bf k},
\end{eqnarray}
where ${\bf x}_0$ is the SZE cluster center,
${\bf x}_1$ is the pointing direction,
$B({\bf x})$ is the primary beam,
and the integration accounts for the frequency band-smearing effect.
These $\{v_p\}$ are equivalent to the band-visibilities at ${\bf x}_1={\bf x}_0$
and therefore are real numbers.
They indicate the cluster profile in the visibility space.

\begin{figure}
\plotone{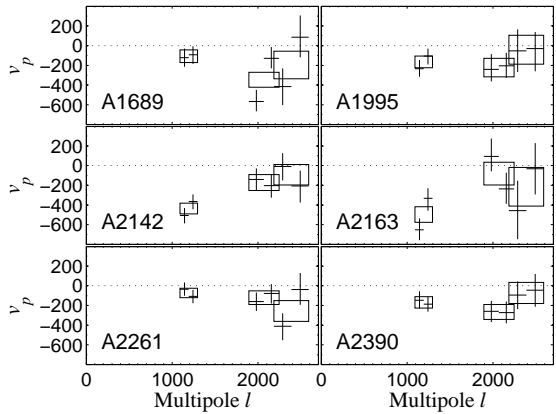}
  \caption{\small
  Visibility profiles $\{v_p\}$ of AMiBA SZE clusters.
  The data for the crosses are summarized in Table~\ref{tab-mcmc},
  while the boxes are the noise weighted average of the two frequency bands at each baseline length.
  \label{fig-vis-profile}}
\end{figure}

Markov Chain Monte-Carlo (MCMC) approach with Metropolis-Hastings sampling
is used for this eight-dimensional likelihood analysis.
For each cluster three MCMC chains of 200,000 samples are used.
The results are given in Table~\ref{tab-mcmc} and Figure~\ref{fig-vis-profile}.
For all clusters the expected SZE decrement in flux density is evident.
We note that although the primary CMB makes a non-negligible contribution
to our observed visibilities \citep{amiba07-liu},
it appears with random phase in $u$-$v$ space and 
thus will not bias our analysis here but rather contribute as part of the error bars.

\begin{table*}
	\begin{center}
	\caption{SZE Centers and Visibility Profiles for AMiBA Clusters\label{tab-mcmc}}
	\scriptsize
	\begin{tabular}{c | c c | c c c c c c}
		\tableline\tableline
		cluster	& \multicolumn{2}{c|}{SZE center ${\bf x}_0$ (J2000)}
						& $v_1$	& $v_2$	& $v_3$	& $v_4$	& $v_5$	& $v_6$\\
						&		RA					& DEC	
						&	(mJy)				&	(mJy)				&	(mJy)				&	(mJy)				&	(mJy)				&	(mJy)\\
		\tableline
		A1689	&	$13^{\rm h}11.41\pm 0.03^{\rm m}$ & $-1^\circ20.7\pm 0.4'$		
					&	$-123\pm92$	&	$-94\pm89$	&	$-566\pm109$	&	$-130\pm108$	&	$-414\pm185$	&	$85\pm212$	\\
		A1995	&	$14^{\rm h}53.12\pm 0.06^{\rm m}$ & $58^\circ02.6\pm 1.0'$		
					&	$-234\pm86$	&	$-106\pm80$	&	$-241\pm138$	&	$-207\pm126$	&	$-53\pm216$	&	$-32\pm200$	\\
		A2142	&	$15^{\rm h}58.30\pm 0.03^{\rm m}$ & $27^\circ13.8\pm 0.5'$
					&	$-508\pm78$	&	$-366\pm76$	&	$-140\pm113$	&	$-205\pm117$	&	$-9\pm139$	&	$-206\pm162$ \\
		A2163	&	$16^{\rm h}15.73\pm 0.04^{\rm m}$ & $-6^\circ09.5\pm 0.5'$
					&	$-652\pm110$	&	$-332\pm113$	&	$93\pm167$	&	$-236\pm158$	&	$-457\pm298$	&	$-31\pm260$	\\
		A2261	&	$17^{\rm h}22.46\pm 0.06^{\rm m}$ & $32^\circ08.9\pm 0.5'$		
					&	$-34\pm68$	&	$-108\pm67$	&	$-160\pm94$	&	$-78\pm99$	&	$-412\pm138$	&	$-37\pm163$	\\
		A2390	&	$21^{\rm h}53.72\pm 0.05^{\rm m}$ & $17^\circ38.8\pm 0.7'$		
					&	$-148\pm87$	&	$-185\pm75$	&	$-260\pm107$	&	$-273\pm112$	&	$-96\pm141$	&	$-44\pm166$	\\
		\tableline
	\end{tabular}
\tablecomments{
The maximum-likelihood results
for the coordinates of SZE cluster centers ${\bf x}_0$ and band-visibility profiles $\{v_p\}$.
The values $v_1$--$v_6$ correspond to the multipole bands of 
$\ell=[1092,1193]$, $[1193,1295]$, $[1891,2067]$, $[2067,2243]$, $[2184,2387]$, $[2387,2590]$ respectively
(see Sec.~\ref{sec-res}, Sec.~\ref{sec-vp}).
$v_1$, $v_3$, $v_5$ correspond to the frequency band of $[86, 94]$~GHz,
and $v_2$, $v_4$, $v_6$ correspond to $[94, 102]$~GHz.
The three baseline lengths $L^b$ are 
$0.606$~m (for $v_1$, $v_2$), $1.05$~m (for $v_3$, $v_4$), and $1.212$~m (for $v_5$, $v_6$).
}
	\end{center}
\end{table*}

We emphasize that these results for $\{v_p\}$ and ${\bf x}_0$ are independent of cluster model.
They can be further compared or combined with other experimental results,
or fitted with a specific cluster model, for further study.
The approach here is different from that in \citet{amiba07-liu},
where an isothermal $\beta$-model is directly fitted with the visibilities
to estimate the central brightness $I_0$,
with the core radius $\theta_{\rm c}$ and the power index $\beta$ taken from X-ray analysis.

\section{Tests for Systematic Error}
\label{sec-sys-err}

To verify that our detections of SZE clusters are real
rather than from the instrument or from the foreground,
we implement several tests.

\subsection{Differencing Maps}

\begin{figure}[!htb] 
	\plotone{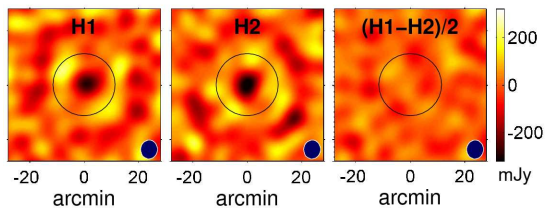}
  \caption{\small  
  Both dirty images (left and middle) of A2142 analyzed from two equally divided sub datasets
  show a clear SZE signal of flux decrement,
	while their difference (right) indicates no recognizable signal but noise.
  \label{fig-a2142-diff}}
\end{figure}

A commonly used powerful test is the so-called sum-and-difference test,
where the temporal data are divided into two subsets of equal size and then processed separately. 
Figure~\ref{fig-a2142-diff} shows an example for A2142,
where the two half-data images (left and middle) show a clear signal
while their difference (right) reveals no signal but noise.
Their average is very close to the overall dirty image shown in Figure~\ref{fig-sz_images}.
We have verified that such feature for the existence of signal is independent of 
the scheme for dividing the data into two halves.
All data of the 6 clusters have passed this test.

\subsection{Other systematic tests}

We carried out several other tests for systematics. Long-duration
(12-hour) observations of planets were used to check the stability of
the system; an independent analysis path produced results for A2142
consistent to high accuracy with those presented here; and several
two-patch blank-sky observations showed no signal above the expected
CMB confusion \citep{amiba07-lin}. The noise properties of the lag
data were investigated in detail indicating no systematics and no
significant non-Gaussianity \citep{amiba07-nishioka}. The asymmetry of
the antenna beams \citep{asym-beam} was shown to be negligible. 
The pointing error is small enough that it
has negligible effects on our results \citep{amiba07-koch-mount}, and
the radio alignment was tuned to improve efficiency and tested for
stability \citep{amiba-wu-radio}.

\section{Scientific Implications}
\label{sec-sci}

The scientific implications of the AMiBA SZE results presented here
are further studied by companion papers.
The Hubble constant is estimated by first deriving the angular diameter distances of the clusters,
and found to be $H_0 = 50^{+16+18}_{-16-23}~{\rm km}\,{\rm s}^{-1}{\rm Mpc}^{-1}$ \citep{amiba07-koch}
in a flat $\Lambda$CDM cosmology with $\Omega_{\rm m}= 0.3$ and $\Omega_\Lambda= 0.7$.
We also estimate the integrated Compton parameter $Y_{2500}$ \citep{amiba07-huang},
which is the Compton $y$ parameter integrated out to the angular radius 
at which the mean overdensity of the cluster is equal to 
2500 times the critical density of the universe at that redshift.
Table~\ref{tab-sz-properties} summarizes the results,
as compared with the results from observations at 
30 GHz \citep[OVRO/BIMA, deduced from][relativistic correction considered]{ovro08} and 
145 GHz \citep[SuZIE~II, deduced from][relativistic correction considered]{suzie204}.
Our results are consistent with those from OVRO/BIMA except for A1995, 
but seem to be systematically lower than those from SuZIE~II.
We also investigate the scaling relations between the $Y_{2500}$ from AMiBA data
and various X-ray derived properties such as the gas temperature $T_{\rm e}$, 
total mass $M_{2500}$ and luminosity $L_{x}$.
The scaling powers of these relations are consistent
with the predictions of the self-similar model \citep{amiba07-huang}.
In the above studies, due to the missing flux problem
the spherical isothermal $\beta$ model is used with the spectral index $\beta$ 
and the core radius estimated by the X-ray data (see Table~\ref{tab-x-properties}),
and the normalization calibrated by the AMiBA SZE visibilities analyzed here \citep{amiba07-liu}. 

\begin{table}
	\begin{center}
	\caption{Comparison of AMiBA SZE Results with Other Observations.\label{tab-sz-properties}}
	\begin{tabular}{c | c c c}
		\tableline\tableline
		Cluster	& \multicolumn{3}{c}{$Y_{2500}$ ($\times 10^{-10}$ sr)}	\\
						&	OVRO/BIMA					&	AMiBA					&	SuZIE~II	\\
		\tableline
		A1689	&	$2.17\pm0.14$	&  $3.13\pm0.96$	&	$4.65^{+0.61}_{-0.51}$\\
		A1995	&	$0.71\pm0.06$	&  $1.60\pm0.36$	&	-			\\
		A2142	&	-							& $14.67\pm2.62$	&	-			\\
		A2163	&	$5.53\pm0.41$	&  $6.32\pm1.10$	&	$5.50^{+0.76}_{-0.70}$\\
		A2261	&	$1.51\pm0.18$	&  $1.36\pm0.71$	&	$4.46^{+1.70}_{-0.94}$\\
		A2390	&	-							&  $1.69\pm0.67$	&	$3.69^{+0.56}_{-0.57}$\\
		\tableline
	\end{tabular}
	\tablecomments{
    The integrated Compton parameter $Y_{2500}$ measured by AMiBA \citep[86--102 GHz;][]{amiba07-huang}
		as compared with results from OVRO/BIMA \citep[30 GHz; deduced from ][relativistic correction considered]{ovro08}
		and SuZIE~II \citep[145 GHz; deduced from ][relativistic correction considered]{suzie204}.}
	\end{center}
\end{table}

We also perform a joint analysis of our SZE data with the weak gravitational lensing data
from Subaru observations \citep{amiba07-umetsu}.
For the four clusters of A1689, A2142, A2261, and A2390, the two data sets are found
to be in great agreement in morphology.
Quantitative analysis even yields an estimation for the baryonic fraction of 
$f_{\rm b}(<r_{200}) = 0.133\pm 0.027$
and concludes that 
when compared with the cosmic baryonic fraction $\Omega_{\rm b}/\Omega_{\rm m}=0.171\pm 0.009$ \citep{dunkley08},
$22\pm 16$~\% of the baryons are missing from the hot phase of clusters \citep{amiba07-umetsu}.
The morphological agreement between our data and the lensing data 
and the consistency of $f_{\rm b}$, $H_0$, $Y_{2500}$ and scaling relations with literature are encouraging, 
indicating that AMiBA is a reliable CMB telescope.

\section{Conclusion}

We successfully detected six SZE clusters with the seven-element AMiBA in its compact configuration.
The analysis method and results presented here mark a milestone for the AMiBA project and provides
the first successful results for SZE clusters in the 3-mm band.
These results are consistent with the published results based on data from other observations,
and provide complementary information at the same time.
A number of tests show that the system performs as expected from its design,
and we anticipate that the current expansion to a 13-element system
\citep{amiba07-ho-mpla,amiba07-wu-mpla} will boost its capability for the study of CMB cosmology.

\acknowledgments

We thank the Ministry of Education, the National Science Council,
the Academia Sinica, and National Taiwan University for their support of this project. 
We thank the Smithsonian Astrophysical Observatory for hosting the AMiBA project staff at the SMA Hilo Base Facility. 
We thank the NOAA for locating the AMiBA project on their site on Mauna Loa. 
We thank the Hawaiian people for allowing astronomers to work on their mountains in order to study the Universe.
This work is also supported by National Center for Theoretical Science, and
Center for Theoretical Sciences, National Taiwan University for J.H.P.~Wu.
We are grateful for computing support from the National Center for High-Performance Computing, Taiwan.
Support from the STFC for M.~Birkinshaw and K.~Lancaster is also acknowledged.



\begin{thebibliography}{}

\bibitem[Allen (2000)]{Allen2000}
  Allen, S.W., 2000, \mnras, 315,269
\bibitem[Benson {\it et al.}(2004)]{suzie204}
  Benson, B.A., {\it et al.}\, 2004, \apj, 617, 829  
\bibitem[Birkinshaw \& Lancaster(2007)]{birkinshaw-lancaster-07}
	Birkinshaw, M., Lancaster, K.\ 2007, NewAR, 51, 346
\bibitem[Bohringer {\it et al.}(1998)]{cite-a2390-beta}
  B\"oehringer, H., {\it et al.}\, 1998, A\&A, 334, 789 
\bibitem[Chen {\it et al.}(2008)]{amiba07-chen}
  Chen, M.-T., {\it et al.}\ 2008, {\apj}S, submitted
\bibitem[Dunkley {\it et al.}(2008)]{dunkley08}
	Dunkley, J., {\it et al.}\ 2008, {\apj}S, submitted (arXiv:astro-ph/0803.0586)
\bibitem[Grego {\it et al.}(2001)]{bima-ovro01}
  Grego, L., {\it et al.}\ 2001, \apj, 552, 2
\bibitem[Halverson {\it et al.}(2008)]{apex-sz08}
	Halverson, N.W., {\it et al.}\ 2008,  \apj, submitted
\bibitem[Ho {\it et al.}(2008a)]{amiba07-ho}
	Ho, P.T.P., {\it et al.}\ 2008a,  \apj, submitted
\bibitem[Ho {\it et al.}(2008b)]{amiba07-ho-mpla}
	Ho, P.T.P., {\it et al.}\ 2008b,  \mpla, 23, 1243
\bibitem[Hogbom (1974)]{miriad-clean}
	Hogbom, J.A., 1974, A\&AS, 15, 417
\bibitem[Huang {\it et al.}(2008)]{amiba07-huang}
	Huang, C.W., {\it et al.}\ 2008, \apj, submitted
\bibitem[Kneissl {\it et al.}(2001)]{Kneissl2001}
	Kneissl, R., {\it et al.}\ 2001, \mnras, 328, 783
\bibitem[Koch {\it et al.}(2006)]{koch06}
	Koch, P., Raffin, P.A., Wu, J.-H.P., {\it et al.}\ 2006, Proc.~EuCAP (ESA SP-626), p.668.1
\bibitem[Koch {\it et al.}(2008a)]{amiba07-koch-mount}
	Koch, P., {\it et al.}\ 2008a, {\apj}S, submitted
\bibitem[Koch {\it et al.}(2008b)]{amiba07-koch}
	Koch, P., {\it et al.}\ 2008b, \apj, submitted
\bibitem[Kosowsky(2003)]{Kosowsky2003}
	Kosowsky, A., {\it et al.}\ 2003, New Astronomy Reviews, 47, 939
\bibitem[LaRoque(2006)]{bima-sz06}
  LaRoque, S., {\it et al.}\ 2006, \apj, 652, 917
\bibitem[Lancaster {\it et al.}(2005)]{vsa05}
  Lancaster, K., {\it et al.}\ 2005, \mnras, 359, 16
\bibitem[Lancaster {\it et al.}(2007)]{ocra07}
	Lancaster, K., {\it et al.}\ 2007, \mnras, 378, 673
\bibitem[Leitch {\it et al.}(2005)]{dasi-05}
	Leitch, E.M., {\it et al.}\ 2005, \apj, 624, 10
\bibitem[Lin {\it et al.}(2004)]{sz04-lin}
	Lin, K.Y., {\it et al.}\ 2004,  \apj, 608, L1
\bibitem[Lin {\it et al.}(2008)]{amiba07-lin}
	Lin, K.Y., {\it et al.}\ 2008,  \apj, submitted
\bibitem[Liu {\it et al.}(2008)]{amiba07-liu}
	Liu, G.C., {\it et al.}\ 2008, \apj, submitted
\bibitem[Markevitch {\it et al.}(1998)]{markevitch98}
	Markevitch, M., Forman, W.~R., Sarazin C.~L., Vikhlinin A., 1998, \apj, 503, 77
\bibitem[Mason {\it et al.}(2001)]{ovro01}
  Mason, B.S., Myers, S.T., \& Readhead, A.C.S., 2001, \apj, 555, L11
\bibitem[MIRIAD-ATNF (2008)]{miriad}
	MIRIAD, Australia Telescope National Facility,\\ http://www.atnf.csiro.au/computing/software/miriad/
\bibitem[Molnar {\it et al.}(2008)]{amiba07-molnar}
	Molnar, S., {\it et al.}\ 2008, \apj, submitted
\bibitem[Morandi {\it et al.}(2008)]{ovro08}
	Morandi, A., {\it et al.}\ 2008, \mnras, in press
\bibitem[Muchovej {\it et al.}(2007)]{sza07}
	Muchovej, S., {\it et al.}\ 2007, \apj, 663, 708
\bibitem[Nishioka {\it et al.}(2008)]{amiba07-nishioka}
	Nishioka, H., {\it et al.}\ 2008, \apj, submitted
\bibitem[Readhead {\it et al.}(2004)]{cbi-04}
	Readhead, A.C.S., {\it et al.}\ 2004, Science, 306, 836
\bibitem[Reese {\it et al.}(2002)]{Reese2002}
	Reese, E. D., Carlstrom, J. E., Joy, M., Mohr, J. J., Grego, L., Holzapfel, W. L., 2002, \apj, 581, 53 
\bibitem[Ruhl(2004)]{Ruhl2004}
	Ruhl, J., 2004, The South Pole Telescope, 11
\bibitem[Sievers {\it et al.}(2005)]{cbi-05}
	Sievers, J.L., {\it et al.}\ 2005, \apj, submitted (astro-ph/0509203)
\bibitem[Sanderson \& Ponman(2003)]{Sanderson2003}
	Sanderson A. J. R., Ponman T. J., 2003, \mnras, 345, 1241
\bibitem[Sunyaev and Zel'dovich (1972)]{s-z}
  Sunyaev, R. A., Zeldovich, Y. B., 1972, Comments on Astrophysics and Space Physics, 4, 173
\bibitem[Umetsu {\it et al.}(2004)]{amiba04-umetsu}
	Umetsu, K., {\it et al.}, 2004, \mpla, 19, 1027
\bibitem[Umetsu {\it et al.}(2008)]{amiba07-umetsu}
	Umetsu, K., {\it et al.}\ 2008, \apj, submitted
\bibitem[Wu {\it et al.}(2001)]{asym-beam}
	Wu, J.-H.P., {\it et al.}\ 2001, \apjs, 132, 1
\bibitem[Wu {\it et al.}(2004)]{amiba04-wu}
	Wu, J.-H.P., {\it et al.}\ 2004, \mpla, 19, 1019
\bibitem[Wu {\it et al.}(2008a)]{amiba07-wu-mpla}
	Wu, J.-H.P., {\it et al.}\ 2008a, \mpla, 23, 1675
\bibitem[Wu {\it et al.}(2008b)]{amiba-wu-radio}
	Wu, J.-H.P., {\it et al.}, 2008b, in preparation


\end{thebibliography}
\end{document}